\documentstyle[12pt]{article}

\begin{document}

\title{$J/\Psi$ suppression in nuclear collisions at SPS
energies\footnote{Supported by BMBF and GSI Darmstadt}}
\author{W. Cassing and C. M. Ko\thanks{Permanent address:
Cyclotron Institute and Physics Department, Texas A\&M University,
College Station, Texas 77843}\\
Institut f\"ur Theoretische Physik, Universit\"at Giessen \\
D-35392 Giessen, Germany}

\maketitle

\begin{abstract}
We have carried out a first calculation of $J/\Psi$ production from
nuclear collisions within the covariant transport approach HSD, which
has been very successful in describing both hadronic and
electromagnetic observables from heavy-ion collisions at intermediate
and high energies. The production of $J/\Psi$'s is based on the Lund
string fragmentation model, while its interactions with hadrons are
included by conventional cascade-type two-body collisions. Adopting 6
mb for the $J/\Psi$-baryon cross sections and 3 mb for the
$J/\Psi$-meson cross sections above the $D\bar{D}$ threshold, we find
that data on $J/\Psi$ suppression from both proton-nucleus and
nucleus-nucleus collisions (including Pb + Pb) can be explained without
assuming the formation of a quark-gluon plasma in these collisions. Our
microscopic studies thus confirm the  suggestion of Gavin {\it et al.}
that $J/\Psi$ suppression observed in nuclear collisions is largely due
to absorption by comovers, i.e., the produced mesons.
\end{abstract}

\vspace{1cm}

\newpage
Because of color screening a $J/\Psi$ dissolves in a quark-gluon plasma.
Matsui and Satz \cite{matsui} thus proposed that a suppression of
the $J/\Psi$ yield in ultra-relativistic heavy-ion collisions is a
plausible signature for the formation of the quark-gluon plasma
in these collisions.  This suggestion has stimulated a number of
heavy-ion experiments at CERN SPS to measure the $J/\Psi$ production via
its dimuon decay. Indeed,
these experiments have shown a significant reduction of the
$J/\Psi$ yield when  going from proton-nucleus to
nucleus-nucleus collisions \cite{na38,na50}. Especially for Pb + Pb at
160 GeV/u an even more dramatic reduction of $J/\Psi$ has been reported
by the NA50 collaboration \cite{gonin}.

To understand the experimental results, models based on $J/\Psi$ 
absorption by hadrons have also been proposed. In Ref. \cite{gerschel}, 
Gerschel and H\"ufner have shown that the observed suppression of 
$J/\Psi$ in nuclear collisions is consistent with the hadronic 
absorption scenario if one assumes that the $J/\Psi$-nucleon absorption 
cross section is about 6-7 mb. On the other hand, Gavin {\it et al.} 
\cite{gavin}, based also on the hadronic absorption model, have found 
that although $J/\Psi$ absorption by nucleons is sufficient to explain 
the proton-nucleus data, it can not account for the suppression seen in 
nucleus-nucleus collisions.  Introducing $J/\Psi$ absorption by the 
produced mesons, which they call the `comovers', with a cross section 
of about 3 mb, they can also obtain a satisfactory account of the 
nucleus-nucleus data.  They attribute the difference between their 
conclusion and that of Ref.  \cite{gerschel} to the different 
consideration in the thickness of nuclear matter where the produced 
$J/\Psi$ has to pass through.  In both studies, the dynamics of the 
collisions is based on the Glauber model, so a detailed space and time 
evolution of the colliding system is not included.  In particular, the 
transverse expansion of the system is ignored in these studies. 
Especially for nucleus-nucleus collisions involving heavier beams, such 
as the recent Pb+Pb collisions at 160 GeV/nucleon, the dynamics is more 
complex than in proton and S induced reactions. Although the hadronic 
absorption model of Ref. \cite{gavin} has been shown to also explain 
the data \cite{gavin1}, an improved study based on a more realistic 
dynamic model will be very useful in clarifying the underlying 
assumptions.

In this work we will carry out a study of $J/\Psi$ production and 
suppression at SPS energies using the covariant transport approach HSD 
\cite{cassing}.  This nonequilibrium model has been shown to describe 
satisfactorily both the measured hadronic observables (rapidity 
distributions, transverse momentum spectra etc.  \cite{cassing}), which 
are sensitive to the final stage collision dynamics, and dilepton 
spectra \cite{cassing1,cassing2,brat96}, which reflect also the initial 
hot dense stage of the collisions.  It thus gives a more realistic 
description of the heavy-ion reaction dynamics than those used in Refs. 
\cite{gerschel,gavin,gavin1}.  Within this approach we can check if the 
simple hadronic absorption model used in both Ref. \cite{gerschel} and 
Refs. \cite{gavin,gavin1} can indeed explain quantitatively the 
observed $J/\Psi$ suppression.  Furthermore, we can check if the 
$J/\Psi$ is destroyed by nucleons before mesons are produced, since 
this assumption has been used in Ref. \cite{blaizot} to argue that 
$J/\Psi$ absorption by mesons should be neglected, and the large 
suppression observed in Pb+Pb collisions might partly be due to the 
quark-gluon plasma.  We will not address the question of whether the 
magnitude of the $J/\Psi$-hadron cross sections used  is correct or can 
be justified by nonperturbative QCD. According to Ref. \cite{kharzeev}, 
these cross sections might be negligibly small in hadronic matter due 
to the small size of the $J/\Psi$ and its large mass gap from open 
charms. However, it might well be true that the $c\bar{c}$ pair is 
first produced in a color-octet state together with a gluon and that 
this more extended configuration has a larger interaction cross section 
with baryons and mesons (i.e., 6 mb and 3 mb, respectively) before the 
$J/\Psi$ singlet state finally emerges.

In HSD a $J/\Psi$ (or better $c\bar{c}$ pair) is produced from string 
fragmentation \cite{LUND} in the initial stage of a nuclear collision. 
Since the probability of producing a $J/\Psi$ is very small, a 
perturbative approach is used in our study:  i.e., whenever an 
$s\bar{s}$ pair ($\Phi$-meson) is produced in the string decay, a 
$J/\Psi$ is produced with a probability factor $W$, which is given by 
the ratio of the $J/\Psi$ to $\Phi$ cross section at a center-of-mass 
energy $\sqrt{s}$ of the baryon-baryon collision, i.e.,
\begin{equation}
W= \frac{\sigma_{BB \rightarrow J/\Psi + X} (\sqrt{s})}
{\sigma_{BB \rightarrow \Phi + X} (\sqrt{s})}.
\end{equation}
We then follow the motion of the
$J/\Psi$ in hadronic matter throughout the collision dynamics by treating
its collisions with hadrons
in the same way as for other hadron-hadron collisions \cite{cassing}.
We use as in Refs. \cite{gavin,gavin1} 6 mb for $J/\Psi$-baryon
collisions ($J/\Psi+B\to\Lambda_c+\bar D$)
and 3 mb for $J/\Psi$-meson collisions ($J/\Psi+m\to D\bar D$) once the
energies are above the threshold for these reactions. Since
the $J/\Psi$ production is treated perturbatively, light hadrons
are not affected by these collisions in their propagation,
however, the $J/\Psi$ is destroyed.

Since in experiment the $J/\Psi$ is measured in nuclear collisions from 
its decay into dimuons, we calculate explicitly the dimuon invariant 
mass spectra from the collisions. This includes not only the decay of 
the $J/\Psi$ but also the decay of other vector mesons ($\rho$, 
$\omega$, and $\phi$) as well as the Dalitz decay of $\pi$, $\eta$, 
$\omega$, etc.  Details on calculating the dilepton spectra from 
heavy-ion collisions up to invariant masses of about 1.5 GeV can be 
found in Refs.  \cite{cassing1,cassing2}.  Since both the Drell-Yan and 
open charm contributions are important for dileptons with invariant 
masses above 1.5 GeV, the latter being known e.g. for p + W reactions 
\cite{carlos}, we have simulated their yield by a background term which 
is fitted to the dimuon yield for p + W at 200 GeV/u.

We have carried out calculations for p+W, S + W and Pb + Pb collisions 
at 200 GeV/u as well as Pb+Pb collisions at 160 GeV/nucleon. All 
results presented below are obtained at an impact parameter of 2 fm.  
In Fig. 1 we show the dilepton invariant mass spectra for the three 
reactions normalized to the number of charged particles in the 
pseudorapidity bin 3.7 $\leq \eta \leq$ 5.2 and compare them with the 
experimental data from Ref.  \cite{helios}. Since there are no data 
from the HELIOS-3 collaboration for Pb+Pb at 200 GeV/u, the S+W data at 
200 GeV/u are employed for the comparison with the calculated results 
for Pb+Pb. It is seen from Fig. 1 that in the p+W and S+W cases the 
theoretical results agree well with the data on an absolute scale which 
implies that apart from the low mass dimuon spectrum - which has been 
analysed in Ref. \cite{cassing2} - also the $J/\Psi$ region is 
described reasonably well.  For S+W in the invariant mass range from 
1.3 GeV $\leq$ m $\leq$ 2.5 GeV we miss about a factor of 2 in the 
dilepton yield, which might be due to the contribution from $\pi 
a_1\to\mu^+\mu^-$ \cite{song} and/or an enhancement of open charm 
channels in the nucleus-nucleus case; these channels are not included 
in the present transport approach and require further analysis.  When 
comparing the theoretical spectrum for Pb+Pb with the data for S+W in 
the $J/\Psi$ mass region (Fig. 1c), we find a drastic suppression as 
compared to the S + W reaction. We, therefore, may already conclude 
that the simple calculations of Refs. \cite{gavin,gavin1} are quite 
reasonable and the observed large suppression of $J/\Psi$ production in 
nuclear collisions (especially Pb + Pb) can be accounted for by 
hadronic absorption.

To understand more clearly the $J/\Psi$ absorption in hadronic matter, 
we show in Fig. 2 the time evolution of the $J/\Psi$ abundance for p + 
W and S + W at 200 GeV/u as well as for Pb + Pb at 160 GeV/u. The solid 
curves show results with absorption by both baryons and mesons while 
the dashed curves only reflect the absorption by baryons. The dotted 
lines show the actual number of $J/\Psi$'s in the simulation without 
including any reabsorption.  We see that $J/\Psi$ absorption by mesons 
is important in both S+W and Pb+Pb collisions as suggested in Refs. 
\cite{gavin,gavin1}; this is in contradiction to the assumptions of 
Ref. \cite{blaizot}. On the other hand, there is no sizeable effect 
from $J/\Psi$-meson collisions for p+W due to the low meson densities 
involved. The enhancement of $J/\Psi$ suppression in Pb + Pb collisions 
compared to S + W reactions is basically due to a longer reaction time 
with comovers as seen from Fig. 2b) and c).

For Pb+Pb collisions, the propagation length L of the $J/\Psi$ in 
nuclear matter estimated by the Glauber theory is found to saturate 
with impact parameter b $\leq$ 9 fm \cite{gonin,bormio}, i.e., L 
$\approx$ 10.5 - 11.5 fm.  Thus when plotting the $J/\Psi$ suppression 
factor versus L, a sudden and dramatic reduction is found 
experimentally \cite{gonin}. As argued by Gavin and Vogt \cite{gavin}, 
this reduction is an artefact of the representation and should become 
smooth as a function of the transverse energy produced which increases 
drastically with decreasing impact parameter. We have investigated the 
latter question in more detail and show in Fig. 3 the calculated 
$J/\Psi$ suppression factor for Pb + Pb at 160 GeV/u as a function of 
the transverse energy produced normalized to the transverse energy 
$E_T^{max}$ for b = 0 fm. The corresponding values for the impact 
parameter are in steps of 1 fm starting from b = 11 fm to b = 1 fm.  
Indeed, as in the analysis by Gavin and Vogt \cite{gavin1} the $J/\Psi$ 
suppression is found to be smooth in the transverse energy and 
approximately agrees with the preliminary data from \cite{gonin}.

In order to explore if the energy density in these reactions might be 
large enough to create a quark-gluon plasma in some region of space and 
time, we show as a function of time (Fig. 4) the volume with energy 
density above 2, 3, and 4 GeV/fm$^3$ for S + W at 200 GeV/u and Pb + Pb 
at 160 GeV/u for a reaction at b = 2 fm. We do not show the result for 
p+W at 200 GeV because the corresponding volumes are zero in the latter 
case.   In these calculations the energy density is computed as
\begin{equation}
E(x) = (\Delta V(x) \gamma(x))^{-1} \{\sum_{baryons \ i \ \epsilon \
\Delta V} \sqrt{p_i^2 + m_i^2}
+  \sum_{mesons \ j \ \epsilon \ \Delta V} \sqrt{p_j^2 + m_j^2} \},
\end{equation}
where all mesons, but only baryons
that have scattered at least once, have been counted.
In eq. (2), $\gamma(x)$ is the Lorentz-factor associated with
the cell $\Delta V(x)$, which is taken to be
1 fm$^3$(\footnote{The results in Fig. 4 are independent of the cell size
$\Delta V(x)$ from 0.15 to 1.5 fm$^3$.}),
in the nucleus-nucleus center of mass.

It is seen that in Pb+Pb collisions there is an appreciable volume of 
high energy density above 3 GeV/fm$^3$ ($\approx 300$ fm$^3$) for time 
scales of a few fm/c, where a quark-gluon plasma (QGP) might be formed 
in the reaction\footnote{The critical energy density for a phase 
transition to the QGP is not accurately known; the value of 3 
GeV/fm$^3$ is chosen for convenience.}. This region of high energy 
density is about 20 \% of the volume of Pb and - from our point of view 
- should not be adequately described by a hadronic transport theory.  
The actual volume of high energy density (above 3 GeV/fm$^3$) for S + W 
is only about 50 fm$^3$, but the volume per participating projectile 
nucleon is roughly the same. Thus central S + W and Pb + Pb collisions 
should show similar features when normalizing by the number of charged 
particles in a more forward rapidity bin. As an example we mention the 
normalized dimuon spectra (within the HELIOS-3 acceptance) for the two 
reactions in Fig. 1b) and 1c), which within 10 - 20 \% are practically 
the same for invariant masses below 2 GeV.

Since in p+W collisions mesons do not contribute substantially to 
$J/\Psi$ absorption and the energy density is not high either, the fact 
that the theoretical results agree with the data can be used to justify 
the use of 6 mb for the $J/\Psi$-baryon cross sections.  Furthermore, 
the agreement between the theoretical results with the S+W data, where 
quark-gluon plasma effects should be of minor importance, indicates 
that the 3 mb used for the $J/\Psi$-meson cross sections is also 
reasonable.

We note, furthermore, that our analysis yields a maximum suppression of 
$J/\Psi$'s at high baryon and meson density, which also directly 
correlates with a high energy density. Thus alternative assumptions 
about $J/\Psi$ suppression, that rise almost linearly with the energy 
density, cannot be ruled out at the present stage.

In conclusion, we have carried out a first microscopic transport study 
of $J/\Psi$ production and absorption in nuclear collisions.  Including 
only absorption by hadrons, we confirm the results of Refs.  
\cite{gavin,gavin1} based on a simple Glauber model that the observed 
suppression in both proton-nucleus and nucleus-nucleus collisions can 
be explained. In particular, the absorption of $J/\Psi$'s by produced 
mesons is important in nucleus-nucleus collisions and especially for Pb 
+ Pb, where the $J/\Psi$-hadron reactions extend to much larger times 
as compared to the S + W reaction.  The scaling of the $c\bar{c}$ pair 
suppression with the geometrical path L according to Glauber theory is 
found to be misleading, since the dependence on the meson density, and 
thus on the transverse energy produced, is smooth.

Since the hadronic $J/\Psi$ reabsorption is proportional to the hadron 
density and thus also approximately proportional to the energy density, 
we presently cannot rule out a possible dissociation of the $c\bar{c}$ 
pair within local QGP droplets, which are expected to be formed at 
least for energy densities above 3 $GeV/fm^3$. It might also happen 
that the $c\bar{c}$ pairs have a finite probability to escape from the 
QGP-phase; such a scenario would be hard to distinguish experimentally 
from the hadronic reabsorption model. At the present stage we can only 
exclude a full $J/\Psi$ dissociation in the QGP-phase, if the critical 
energy density for the phase transition is below about 1.5 $GeV/fm^3$, 
because the suppression factors calculated for S + W at 200 GeV/u and 
Pb + Pb at 160 GeV/u are then larger than those seen experimentally so 
far. In order to discriminate between the different model assumptions 
we need better information on the $c\bar{c}$ cross sections with 
hadrons; a task we here have to delay for future work.

\vskip 1cm
The authors like to thank E. L. Bratkovskaya and C. Greiner for 
valuable comments and a critical reading of the manuscript. C.M.K. was 
also supported in part by the National Science Foundation under Grant 
No. PHY-9509266 and the Alexander von Humboldt Foundation.

\newpage

\section*{ Figure Captions}

\noindent{{\bf Fig. 1:} Dimuon invariant mass spectra from p+W, S+W, and
Pb+Pb collisions at 200 GeV/nucleon in comparison to the data of the
HELIOS-3 collaboration \cite{helios}. For the case of Pb + Pb (c) we
compare to the data for S+W since no equivalent experimental spectra are
available.}

\vspace{1cm}
\noindent{{\bf Fig. 2:} Time dependence of the $J/\Psi$ abundance for
p+W, S+W at 200 GeV/u and Pb+Pb collisions at 160 GeV/u.
The dotted lines show the
total number of produced $J/\Psi$'s without reabsorption;
the dashed lines display the
$J/\Psi$ number when including absorption by baryons while the full
lines show the results of our calculation when including absorption by both
baryons and mesons as described in the text. The $J/\Psi$ number
calculated is proportional to the actual multiplicity for S + W and
Pb + Pb, whereas the statistics have been increased for p + W by a factor 100.}

\vspace{1cm}
\noindent{{\bf Fig. 3:}  The $J/\Psi$ suppression factor for Pb + Pb at 160
GeV/u as a function of the transverse energy normalized to the maximum
transverse energy $E_T^{max}$ at b = 0 fm. The individual dots stand for
a fixed impact parameter b which decreases in steps of 1 fm from 11 fm to
1 fm.}

\vspace{1cm}
\noindent{{\bf Fig. 4:} Time evolution of the reaction volume with an
energy density above 2, 3, and 4 GeV/fm$^3$ for S+W at 200 GeV/u and
Pb + Pb at 160 GeV/u.}

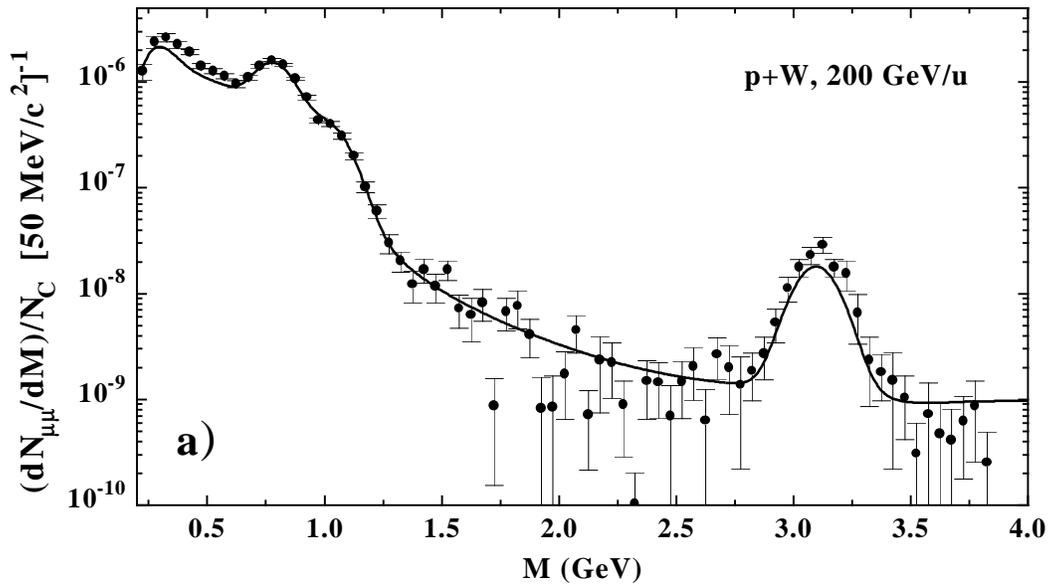
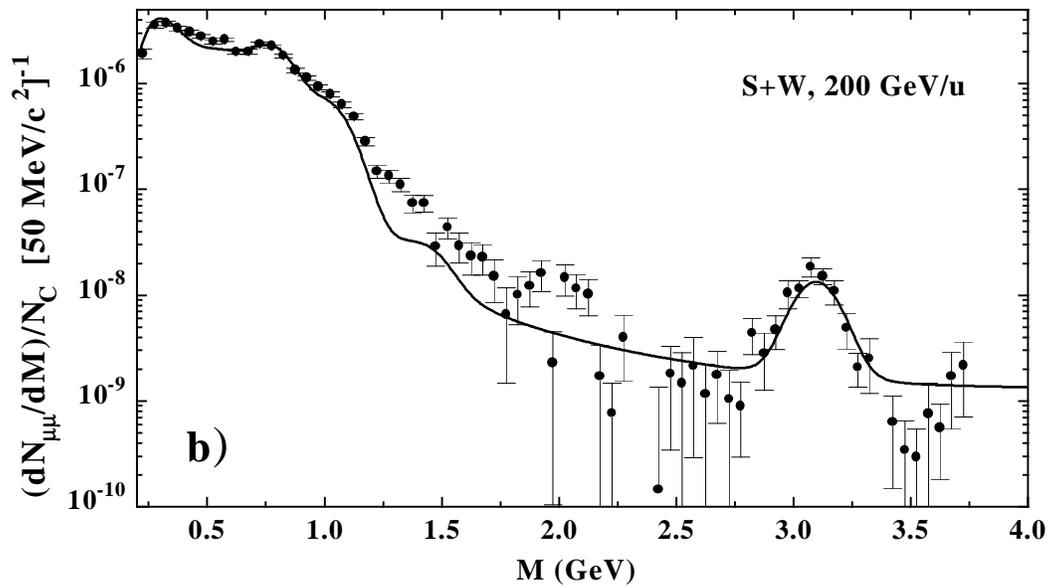
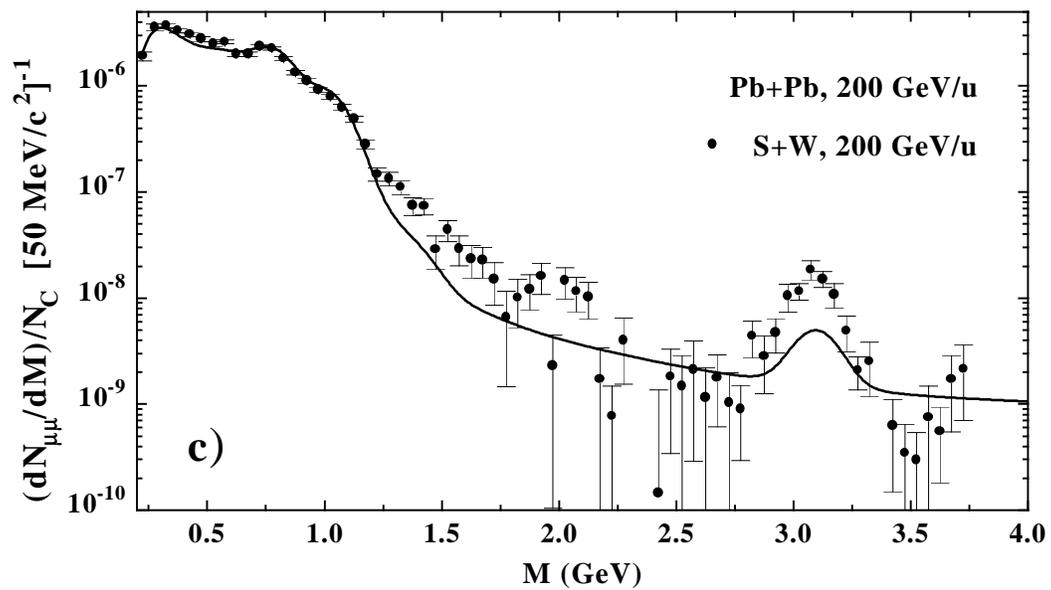
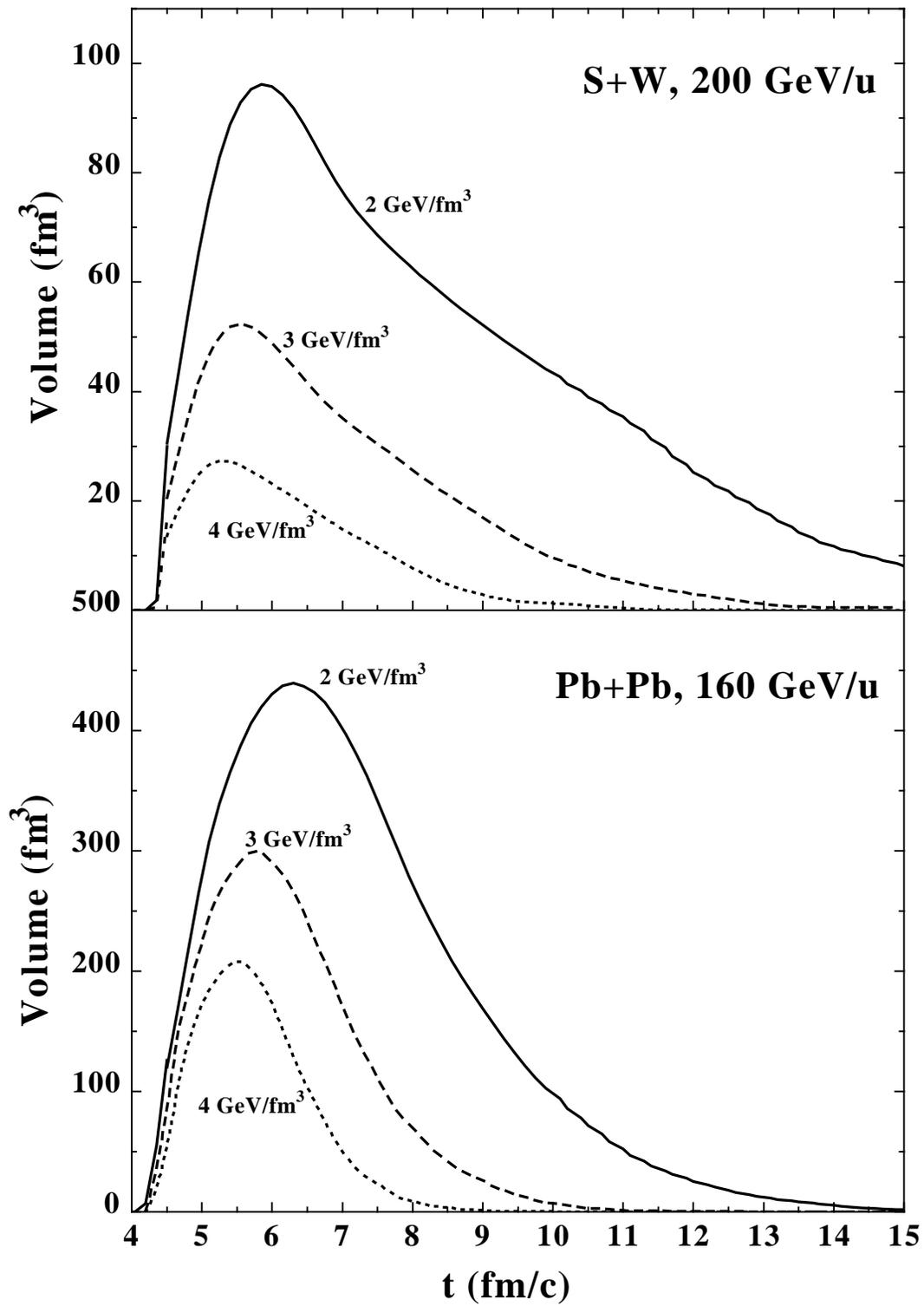
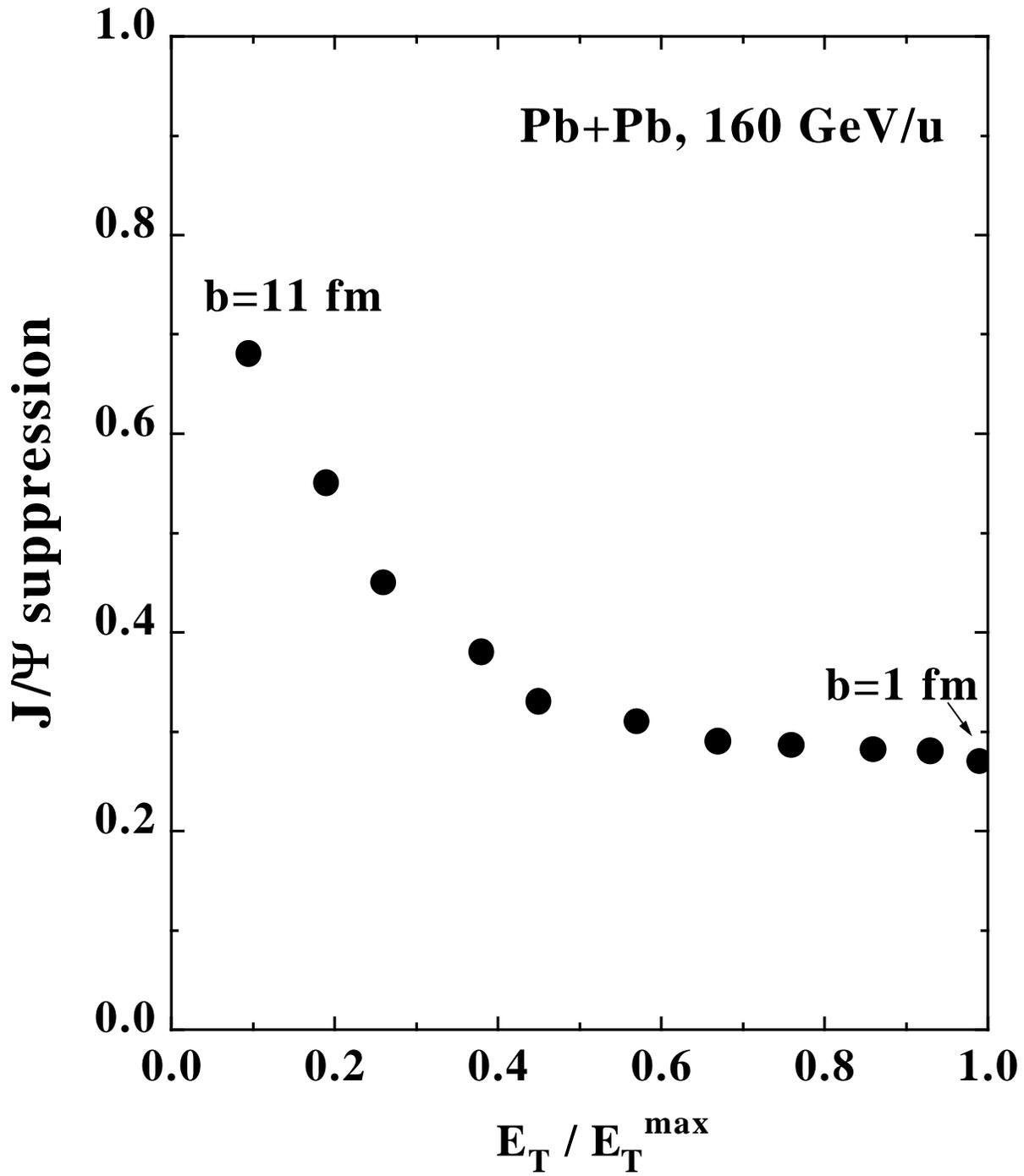
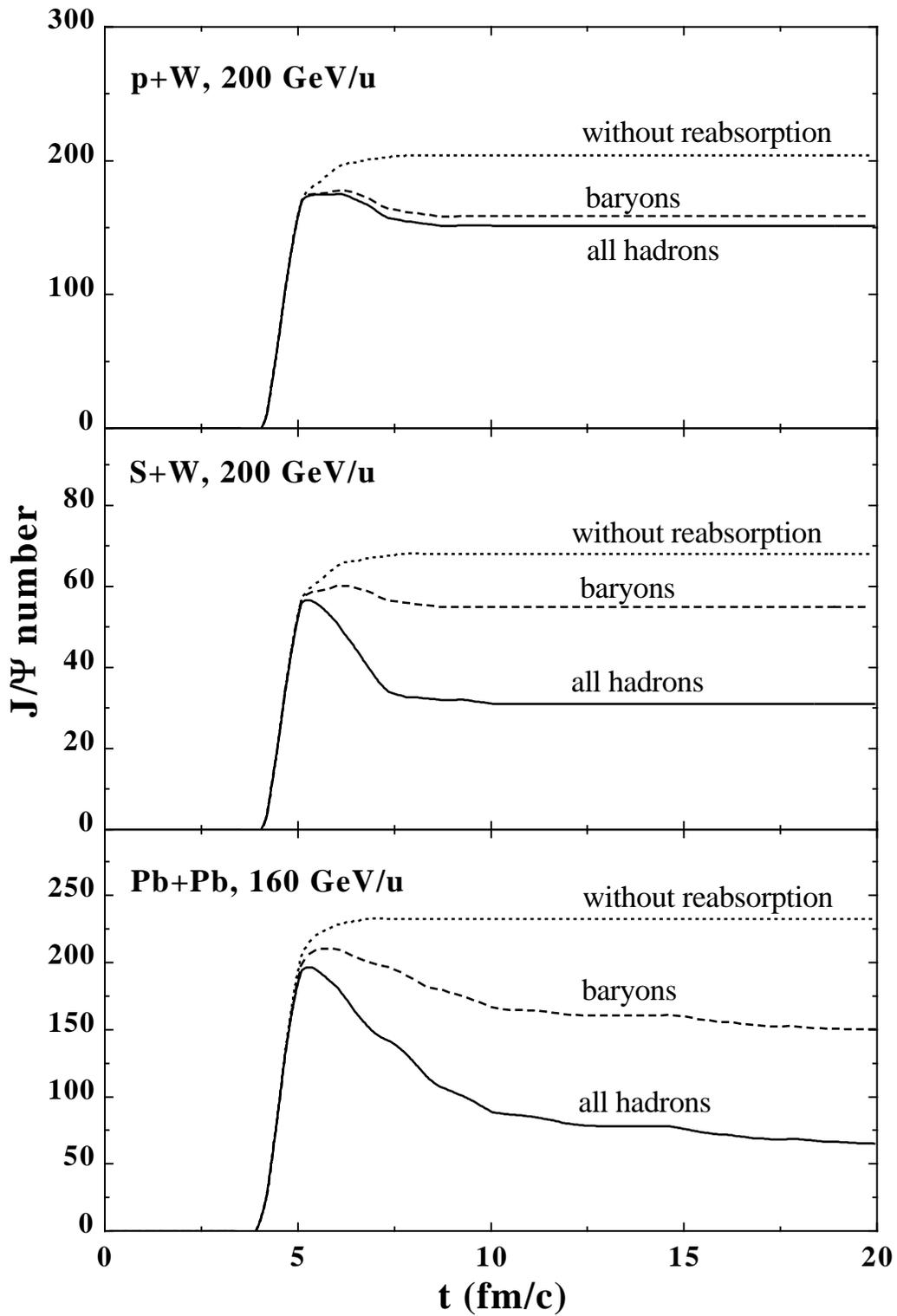
\end{document}